\documentclass[prb,aps,twocolumn,floats]{revtex4-2}
\usepackage{graphicx}
\usepackage{amsmath}

\begin{document}

\title{Topological Green function of interacting systems}
\author{Minh-Tien Tran$^1$, Duong-Bo Nguyen$^2$, Hong-Son Nguyen$^3$, and Thanh-Mai Thi Tran$^1$}
\affiliation{$^1$Institute of Physics, Vietnam Academy of Science and Technology,
Hanoi 10000, Vietnam \\
$^2$Graduate University of Science and Technoly, Vietnam Academy of Science and Technology,
Hanoi 10000, Vietnam \\
$^3$Department of Occupational Safety and Health, Trade Union University,  Hanoi 10000, Vietnam
}

\begin{abstract}
We construct a Green function, which can identify the topological nature of interacting systems. It is equivalent to the single-particle Green function of effective non-interacting particles, the Bloch Hamiltonian of which is given by the inverse of the full Green function of the original interacting particles at zero frequency. The topological nature of the interacting insulators is originated from the coincidence of the poles and the zeros of the diagonal elements of the constructed Green function. The cross of the zeros in the momentum space
closely relates to the topological nature of insulators.
As a demonstration, using the zero's cross, we identify the topological phases of magnetic insulators, where both the ionic potential and the spin exchange between conduction electrons and magnetic moments are present together with the spin-orbital coupling. The topological phase identification is consistent with the topological invariant of the magnetic insulators.
We also found an antiferromagnetic state with topologically breaking of the spin symmetry, where electrons with one spin orientation are in topological insulating state, while electrons with the opposite spin orientation are in topologically trivial one.
\end{abstract}

\maketitle

\section{Introduction}
The topological nature of interacting systems is a fascinating problem. The long-range ordering established by the interaction may intriguingly impact on the topology of the ground state, and as a result exotic states may emerge \cite{Assaad,Rachel}.
The interplay between the topology and the interaction is quite complicated to solve.
One of most fundamental methods for solving the many-body problem of interacting particles is the Green function method \cite{Abrikosov}.
It has widely been used and has many successful applications, ranging from high-energy physics to condensed matters. The single-particle dynamics of interacting systems is closely determined by the (single-particle) Green function.
One of the key characteristics of the Green function is the poles. The poles of the Green function characterize the energy and the life time of the elementary excitations of the interacting systems. The order parameter of a long-range ordering can also be determined by the Green function.
Since the topological insulating states were discovered, there are several proposals to express the topological invariant in terms of the Green function of interacting systems \cite{Volovik,Wang1,Xie,Xie1,Gurarie,Wang,Wang2,Wang3}. However, the general formula for the topological invariant is rather complicated \cite{Volovik,Wang1,Xie,Xie1}. It can be simplified by using the Green function at zero frequency \cite{Wang,Wang2,Wang3}. In interacting insulators the topological invariant can still be determined by an effective Bloch Hamiltonian in the same way as by the non-interacting Bloch Hamiltonian in the non-interacting systems. The effective Bloch Hamiltonian is equivalent to the inverse of the Green function of interacting particles at zero frequency \cite{Wang,Wang2,Wang3}. It has widely been used to identify the topological nature of interacting insulators \cite{Assaad,Tien4,Son,Tien,Mai,Kawakami,Valenti}.
The effective Bloch Hamiltonian can also identify and characterize the topological state of  interacting gapless systems \cite{Knap}.
On the other hand, in the non-interacting systems, the non-zero topological invariant must be accompanied by the existence of the zero
points or vortices
of the Bloch wave function \cite{Hatsugai0,Hatsugai}. In interacting insulators, the vortices are also necessarily occur with the non-zero topological invariant. Therefore, the vortex existence can also be used to identify the topological nature.

Recently, a fundamental identity between the eigenvectors and the eigenvalues of Hermitian operator is rediscovered \cite{Denton}. Using the eigenvector-eigenvalue identity, Misawa and Yamaji showed a close relation between the zero points of the Bloch wave function and the cross of the zeros of the diagonal Green function in the non-interacting systems \cite{Yamaji}. Because the zero points of the Bloch wave function is the topological origin of the systems, they demonstrated that the cross of the zeros of the diagonal Green function is a simple and useful tool for identifying the topological nature of the ground state \cite{Yamaji}. However, the approach is limited to use for the non-interacting systems. Alternatively, the zeros of a projected Green function were used as a local signature of topology \cite{Balents}.

The aim of the present work is twofold. First, we extend the approach, proposed by Misawa and Yamaji, to interacting systems. We construct a Green function, which can determine the topological nature of interacting systems. The coincidence of the poles and the zeros of the constructed Green function is the origin of the vortices of the interacting systems. The cross of the zeros of the diagonal elements of the constructed Green function can identify the topological nature of interacting systems. The second aim of the present work is to investigate the topological nature of  magnetic insulators with the spin-orbital coupling (SOC). The magnetic topological insulators (MTIs) have attracted a lot of research attention due to the emergence of non-trivial topology and magnetism as well as the high potential of their applications in science and technology \cite{Weng,Tokura}. The spontaneous magnetization can play like an external magnetic field, and it can give rise to the quantum anomalous Hall effect \cite{Weng,Tokura}. In the previous studies, we proposed a minimal model for the MTIs \cite{Tien,Mai}. It is based on the Kane-Mele and the double exchange models \cite{KaneMele,Zener}. In the proposed model, the SOC causes the topologically non-trivial insulating state, while the spin exchange between conduction electrons and magnetic moments gives rise to a magnetic ordering \cite{Tien,Mai}. The interplay between the SOC and the spin exchange results in the coexistence of non-trivial topology and magnetism. In particular, the quantum spin Hall (QSH) effect is observed in the antiferromagnetic state \cite{Tien,Mai}. However, in the previous studies the ionic potential was not considered \cite{Tien,Mai}. The ionic potential is originally present in the Kane-Mele model and breaks the sublattice symmetry \cite{KaneMele}. It drives the topological insulator to topologically trivial one \cite{KaneMele}. Both the spin exchange and the ionic potential preserve the topological symmetry between the spin orientations of electrons, i.e., electrons of both spin orientations simultaneously form either topological or topologically trivial insulators
\cite{Tien,KaneMele}. However,
we find that the mutual interplay between the ionic potential and the spin exchange can give rise to a topologically breaking of the spin symmetry. Electrons with one spin orientation form a topological insulator, while electrons with the opposite spin orientation are in topologically trivial insulator.
As a demonstration of the proposed method, we identify the topological nature of the ground state by the cross of the zeros of the proposed Green function. The topological phase identification is consistent with the topological invariant of interacting systems.

The present paper is organized as follows. In Sec. II we construct a single-particle Green function, which can describe the topological nature of interacting systems. In Sec. III  we present a demonstration of using the topological Green function to identify the topological states of magnetic insulators.
Finally, the conclusions are presented in Sec. IV.

\section{Topological Green function}
We consider a general many-body interacting fermion system, Hamiltonian of which can be separated into non-interacting and interacting parts, i.e.,
\begin{equation}
H = H_0 + H_1,
\end{equation}
where $H$ is Hamiltonian of the system, and  $H_0$ ($H_1$) is its non-interacting (interacting) part. The non-interacting Hamiltonian is supposed to be quadratic, i.e.,
\begin{equation}
H_0 = \sum_{i\alpha,j\beta} c^\dagger_{i\alpha} h_{i\alpha,j\beta} c_{j\beta},
\end{equation}
where $c^\dagger_{i\alpha}$ ($c_{i\alpha}$) is the creation (annihilation) operator of fermion at lattice site $i$ with quantum index $\alpha$.
$\alpha$ may include the spin, orbital, sublattice... indices. When the periodic boundary conditions are imposed, the non-interacting Hamiltonian can be rewritten in the momentum space
\begin{equation}
H_0 = \sum_{\mathbf{k},\alpha\beta}  c^\dagger_{\mathbf{k}\alpha} h_{\alpha\beta}(\mathbf{k}) c_{\mathbf{k}\beta}.
\end{equation}
$\hat{h}(\mathbf{k})$ is the so-called Bloch Hamiltonian (we use the hat symbol to denote the matrix form). The fermion dynamics can be analyzed through the Green function.
We consider the Matsubara Green function
\begin{equation}
G_{i\alpha,j\beta}(i\omega)=- \int d\tau \langle \mathcal{T} c_{i\alpha}(\tau) c^\dagger_{j\beta} \rangle e^{i \omega \tau} ,
\end{equation}
where $i\omega$ is the (imaginary) Matsubara frequency. In the zero-temperature limit the discrete Matsubara frequency becomes continuous. The Matsubara Green function can analytically be continued in the complex frequency space.
The non-interacting Green function can be represented through the eigenvectors and eigenvalues of the Bloch Hamiltonian
\begin{eqnarray}
g_{i\alpha,j\beta}(i\omega) = \bigg[ \frac{1}{i\omega \hat{1} - \hat{h}}\bigg]_{i\alpha,j\beta} =
\sum_{n} \frac{\psi_{n}^{(i\alpha)} [\psi_{n}^{(j\beta)}]^{*}}{i\omega-e_n} ,
\label{gf0r}
\end{eqnarray}
where $\psi_{n}^{(j\beta)}$ is the $(j\beta)$-th component of the eigenvector $\hat{\psi}_{n}$ of $\hat{h}$ and $e_n$ is its corresponding eigenvalue, i.e.
\begin{eqnarray}
\hat{h} \hat{\psi}_n = e_n \hat{\psi}_n .
\end{eqnarray}
In the momentum space, the non-interacting Green function reads
\begin{eqnarray}
g_{\alpha\beta}(\mathbf{k},i\omega) &=& \bigg[ \frac{1}{i\omega \hat{1} - \hat{h}(\mathbf{k})}\bigg]_{\alpha\beta} \nonumber \\
&=&
\sum_{n} \frac{\psi_{n}^{(\alpha)}(\mathbf{k}) [\psi_{n}^{(\beta)}(\mathbf{k})]^{*}}{i\omega-e_n(\mathbf{k})} ,
\label{gf0}
\end{eqnarray}
where $\hat{\psi}_{n}(\mathbf{k})$ and $e_n(\mathbf{k})$ are the eigenvector and eigenvalue of $\hat{h}(\mathbf{k})$
\begin{eqnarray}
\hat{h}(\mathbf{k}) \hat{\psi}_n(\mathbf{k}) = e_n(\mathbf{k}) \hat{\psi}_n(\mathbf{k}) .
\end{eqnarray}
The formula in Eq. (\ref{gf0r}) or in Eq. (\ref{gf0}) is just the single-particle representation of the non-interacting Green function. Although the system is many body, the dynamics of non-interacting fermions is equivalent to the single-particle one.

When the interaction is present, the Green function satisfies the Dyson equation
\begin{equation}
\hat{G}(i\omega) = \hat{g}(i\omega) + \hat{g}(i\omega) \hat{\Sigma}(i\omega) \hat{G}(i\omega),
\end{equation}
where $\hat{\Sigma}(i\omega)$ is the self energy. The self energy contains all contributions of interaction to the Green function. The Dyson equation can be rewritten in the real space
\begin{eqnarray}
\sum_{l\gamma}[i\omega \delta_{il}\delta_{\alpha\gamma}-h_{i\alpha,l\gamma}] G_{l\gamma,j\beta}(i\omega) && \nonumber \\
 -\sum_{l\gamma}{\Sigma}_{i\alpha,l\gamma}(i\omega) G_{l\gamma,j\beta}(i\omega)
&=&\delta_{ij}\delta_{\alpha\beta} .
\label{dyson}
\end{eqnarray}
We represent the full Green function in the single-particle form
\begin{equation}
G_{i\alpha,j\beta}(i\omega)=\sum_{n} \frac{\Psi_{n}^{(i\alpha)} [\Psi_{n}^{(j\beta)}]^{*}}{i\omega-E_n},
\label{gf}
\end{equation}
where ${\hat{\Psi}_{n}}$ are required to be orthonormal, i.e.
\begin{equation}
\hat{\Psi}_{n}^{\dagger} \hat{\Psi}_{m} = \sum_{i\alpha} [\Psi_{n}^{(i\alpha)}]^* \Psi_{m}^{(i\alpha)} =\delta_{nm} .
\label{ortho}
\end{equation}
In contrast to the non-interacting case, both $\hat{\Psi}_{n}$ and $E_n$ are unknown yet. Inserting the single-particle representation (\ref{gf}) into the Dyson equation (\ref{dyson}), then multiplying its both sides with $\Psi_m^{(j\beta)}$ and summing over $(j\beta)$ indices, we obtain
\begin{eqnarray}
\sum_{l\gamma} [ h_{i\alpha,l\gamma} + \Sigma_{i\alpha,l\gamma}(i\omega) ]\Psi_{m}^{(l\gamma)}
= E_m \Psi_{m}^{(i\alpha)} .
\label{quasi}
\end{eqnarray}
It turns out that $\hat{\Psi}_m$ and $E_m$ are just the eigenvector and its corresponding eigenvalue of $[\hat{h}+\hat{\Sigma}(i\omega)]$.
One can notice that Eq. (\ref{quasi}) is valid for any frequency $\omega$, and in general, both $\hat{\Psi}_m$ and $E_m$ depend on $\omega$ .
However, Eq. (\ref{quasi}) does not always have solutions satisfying the orthonormality condition for any frequency. Therefore, the representation in Eq. (\ref{gf}) does not always exist. The orthonormality condition in Eq. (\ref{ortho}) is essential, because without it Eq. (\ref{quasi}) cannot be established.
If we set $i\omega=E_m$ in the self energy, Eq. (\ref{quasi}) becomes the so-called quasiparticle equation \cite{Hedin}. In this case, $E_m$ is  the pole of the Green function. In general, the pole $E_m$ is complex, therefore $\hat{\Sigma}(E_m)$ may be non-Hermitian and non-normal. When $\hat{\Sigma}(E_m)$ is non-normal, Eq. (\ref{quasi}) does not guarantee that the eigenvectors $\hat{\Psi}_m$ satisfy the orthonormality condition in Eq. (\ref{ortho}), and the single-particle representation in Eq. (\ref{gf}) may not exist. If we take the limit $\omega \rightarrow 0$ in the self energy, Eq. (\ref{quasi}) becomes
\begin{eqnarray}
\sum_{l\gamma} [ h_{i\alpha,l\gamma} + \Sigma_{i\alpha,l\gamma}(i0) ]\Psi_{m}^{(l\gamma)}
= E_m \Psi_{m}^{(i\alpha)} .
\label{topo}
\end{eqnarray}
In contrast to the quasiparticle case, $\hat{\Sigma}(i0)$ is always Hermitian \cite{Wang,Wang2,Wang3}. Therefore the eigenvalues $E_m$ are real and the eigenvectors $\hat{\Psi}_m$ always exist and can satisfy the orthonormality condition. Indeed,
from the Lehmann representation, one can show that the self energy $\hat{\Sigma}(i\omega)$ is Hermitian only at $\omega=0$ \cite{Wang}.
Only at zero frequency,
$\hat{H}^{\text{topo}} \equiv \hat{h} + \hat{\Sigma}(i0)$ can be considered as an effective  Hamiltonian. For other frequency $[\hat{h} + \hat{\Sigma}(i\omega)]$ may be not Hermitian.
It was proved that the topological Hamiltonian $\hat{H}^{\text{topo}}$ can determine the topological nature of interacting systems \cite{Wang,Wang2,Wang3}.

In the momentum space, Eq. (\ref{topo}) reads
\begin{equation}
[\hat{h}(\mathbf{k}) + \hat{\Sigma}(\mathbf{k},i0)] \hat{\Psi}_{m}(\mathbf{k})
= E_m(\mathbf{k}) \hat{\Psi}_{m}(\mathbf{k}) .
\label{topom}
\end{equation}
The Green function, defined by the single-particle representation in Eq. (\ref{gf}) with eigenvectors $\hat{\Psi}_{m}(\mathbf{k})$ and eigenvalues $E_m(\mathbf{k})$ of $\hat{H}^{\text{topo}}(\mathbf{k})$ can be rewritten as
\begin{equation}
\hat{G}^{\text{topo}}(\mathbf{k},z)=\sum_{m} \frac{\hat{\Psi}_{m}(\mathbf{k}) \hat{\Psi}_{m}^\dagger(\mathbf{k})}{z-E_m(\mathbf{k})}=\frac{1}{z \hat{1} - \hat{H}^{\text{topo}}(\mathbf{k})} .
\label{gftopo}
\end{equation}
$\hat{G}^{\text{topo}}(\mathbf{k},z)$ is called "topological Green function" \cite{Wang2,Knap}. It can describe the topological nature of interacting systems. $\hat{G}^{\text{topo}}(\mathbf{k},z)$ is equivalent to the single-particle Green function of effective non-interacting fermions, the Bloch Hamiltonian of which is given by $\hat{H}^{\text{topo}}(\mathbf{k}) = - [\hat{G}(\mathbf{k},i0)]^{-1}$.

The topological nature of two-dimensional insulators is determined by the Chern number $C_1$, which is the topological invariant of the fiber bundle composed of the Bloch wave functions over the torus of the Brillouin zone. In the physical meaning, the Chern number is just the Hall conductivity $\sigma_{xy}=(e^2/h) C_1$ in unit $e^2/h$. For interacting insulators the Chern number can still be calculated by the generalized Thouless-Kohmoto-Nightingale-den Nijs (TKNN) formula
\begin{eqnarray}
C_1 =  \frac{1}{2\pi} \int d^2 k \mathcal{F}_{xy}(\mathbf{k}) , \label{chern}
\end{eqnarray}
where $\mathcal{F}_{ab}(\mathbf{k})=\partial_{k_{a}} \mathcal{A}_{b}(\mathbf{k}) - \partial_{k_{b}} \mathcal{A}_{a}(\mathbf{k})$,
$\mathcal{A}_{a}(\mathbf{k})=i \sum_{m: E_m<0} \hat{\Psi}_m^\dagger(\mathbf{k}) \partial_{k_{a}} \hat{\Psi}_m(\mathbf{k})$,
$\hat{\Psi}_m(\mathbf{k})$ and $E_m(\mathbf{k})$ are the eigenvector and the eigenvalue of $\hat{H}^{\text{topo}}(\mathbf{k})$ \cite{Wang,TKNN}.
$\hat{\Psi}_m(\mathbf{k})$ can be considered as the Bloch wave function of effective non-interacting fermions,
the Bloch Hamiltonian of which is given by $\hat{H}^{\text{topo}}(\mathbf{k})$. Because the Green function $\hat{G}^{\text{topo}}(\mathbf{k},z)$ describes the dynamics of the  effective non-interacting fermions, their Hall conductivity is still determined by the TKNN formula in
Eq. (\ref{chern}), and it is exactly the Hall conductivity of the original interacting fermions in the insulating state \cite{Wang,Wang2,Wang3}. Therefore the topological nature of insulators is determined by the effective Bloch Hamiltonian $\hat{H}^{\text{topo}}(\mathbf{k})$.
When $C_1=0$ the system is topologically trivial, and non-zero integer $C_1$ indicates the topological insulator.
The topological Hamiltonian $\hat{H}^{\text{topo}}(\mathbf{k})$ can also determine the topological invariants for interacting insulators in higher dimensions \cite{Assaad,Wang}.

One can notice that in the non-interacting case, the Green function, the quasiparticle Green function and the topological Green function are identical. In some cases, where the self energy is independent on frequency, for example in the Hartree-Fock approximation, these Green functions are also identical. However, in general, they are different when the interaction is present.
The topological Green function has the same single-particle form like the non-interacting one and the topological Hamiltonian is Hermitian, therefore its properties are also similar to the ones of the non-interacting Green function \cite{Yamaji}.
One notable property of Hermitian operator is the eigenvector-eigenvalue identity \cite{Denton,Yamaji}. For self containing of the present work, we repeat here the derivation of the eigenvector-eigenvalue identity for the topological Green function, which was derived for the non-interacting systems in Ref. \onlinecite{Yamaji}.
Using the Cramer rule, the diagonal component of the topological Green function can be expressed through its poles and zeros as following
\begin{eqnarray}
G^{\text{topo}}_{\alpha\alpha}(\mathbf{k},z)&\equiv& \sum_{n} \frac{|\Psi_{n}^{(\alpha)}(\mathbf{k})|^2}{z-E_n(\mathbf{k})}
=\frac{\det[z \hat{1}-\hat{M}^{(\alpha)}(\mathbf{k})]}{\det[z \hat{1}-\hat{H}^{\text{topo}}(\mathbf{k})]} \nonumber \\
&=&\frac{\prod_{l}[z-\mathcal{E}_l^{(\alpha)}(\mathbf{k})]}
{\prod_{l}[z-E_l(\mathbf{k})]}
,
\label{diag}
\end{eqnarray}
where $\hat{M}^{(\alpha)}(\mathbf{k})$ is the minor of $\hat{H}_{\text{topo}}(\mathbf{k})$, which is generated by removing the $\alpha$th row and column of $\hat{H}^{\text{topo}}(\mathbf{k})$, $\mathcal{E}_l^{(\alpha)}(\mathbf{k})$ is the $l$th eigenvalue of the minor $\hat{M}^{(\alpha)}(\mathbf{k})$ \cite{Denton,Yamaji}. The diagonal topological Green function has the poles $E_l(\mathbf{k})$, which are the eigenvalues of $\hat{H}^{\text{topo}}(\mathbf{k})$, and the zeros $\mathcal{E}_l^{(\alpha)}(\mathbf{k})$, which are the eigenvalues of the minor $\hat{M}^{(\alpha)}(\mathbf{k})$.
Applying the residue theorem to Eq. (\ref{diag}), one can obtain
\begin{eqnarray}
|\Psi_{n}^{(\alpha)}(\mathbf{k})|^2 = \frac{\prod_{l}[E_n(\mathbf{k})-\mathcal{E}_l^{(\alpha)}(\mathbf{k})]}
{\prod_{l\neq n}[E_n(\mathbf{k})-E_l(\mathbf{k})]} .
\label{eei}
\end{eqnarray}
The relation in Eq. (\ref{eei}) is called "the eigenvector-eigenvalue identity" \cite{Denton,Yamaji}.
It relates the eigenvectors and the eigenvalues of Hermitian operator.
The eigenvector-eigenvalue identity in Eq. (\ref{eei}) shows when $|\Psi_{n}^{(\alpha)}(\mathbf{k})|^2=0$, the pole and the zero of the diagonal topological Green function coincide and vice versa. There might be some momenta at which $|\Psi_{n}^{(\alpha)}(\mathbf{k})|^2$ vanishes.
These are the zero points or vortices of the Bloch wave function \cite{Hatsugai0,Hatsugai}. They give non-zero contributions to the Chern number, and are just the charge of the vortices \cite{Hatsugai0,Hatsugai}.
Therefore, the topological nature of insulators is originated from the vanishing of $|\Psi_{n}^{(\alpha)}(\mathbf{k})|^2$ at some momenta,
or equivalently, from
the coincidence of the pole and the zero of the diagonal topological Green function.
When the pole and the zero coincide, it guarantees a cross of the zeros in the momentum space \cite{Yamaji}. This allows us to use the zero's cross to identify the topological nature of insulators.
The topological Green function for interacting systems was previously noticed \cite{Gurarie,Wang2,Knap}. However, in contrast to the previous studies, only its diagonal elements are relevant to the topological origin through the coincidence their poles and zeros \cite{Yamaji}.

The topological Green function can also describe the topological origin of interacting gapless systems \cite{Knap}. In the gapless systems such as the Weyl or Dirac semimetals, the energy bands touch or cross at the gapless points. The topological gapless points of interacting semimetals are also determined by the touch points of the energy bands of the topological Hamiltonian \cite{Knap}.
From the Cauchy interlacing inequalities \cite{Yamaji}
\begin{eqnarray}
E_l(\mathbf{k}) \leq \mathcal{E}^{(\alpha)}_l(\mathbf{k}) \leq E_{l+1}(\mathbf{k}) ,
\label{interlacing}
\end{eqnarray}
one can notice that the zeros also touch at the gapless points.
Therefore, we can also detect the gapless points by the zeros. However, in addition to the gapless points, the zeros may accidentally touch or cross at high symmetry points of the Brillouin zone. This additionally requires a careful selection of the gapless points from the zero's behaviour. However, if the zeros do not touch or cross, the interacting systems are definitely not gapless.

The quasiparticle Green function also has the single-particle form like the topological Green function, but its effective Hamiltonian is generally not Hermitian. Therefore its eigenvector-eigenvalue identity may not be valid, and the coincidence of its poles and zeros if it exists does not guarantee the existence of vortices and their associate charges. However, it seems that the quasiparticle Green function can also determine the topological invariant of open systems \cite{Hofstetter}.
The vortices or the zero points of the Bloch wave function deeply relate to the topological nature, and they can simply be generalized in any space dimension \cite{Yamaji}.
In the next section, we demonstrate that the topological Green function is a useful tool to identify the topological nature of magnetic insulators.

\section{Topological phases in magnetic insulators}
We consider a minimal model for MTIs \cite{Tien,Mai}.
It consists of a tight binding model of electrons in the presence of the intrinsic SOC and magnetic impurities. The intrinsic SOC is responsible for topological insulating state. Magnetism occurs as a consequence of the spin exchange between electrons and magnetic impurities.
The model Hamiltonian reads
\begin{eqnarray}
H &=&-t\sum\limits_{\langle i,j \rangle, \sigma }c_{i\sigma }^{\dagger }c_{j\sigma}
-i\lambda \sum\limits_{\left\langle \left\langle i,j\right\rangle\right\rangle ,s,s^{\prime }}
\nu _{ij}c_{is}^{\dagger }\sigma _{ss^{\prime}}^{z}c_{js^{\prime }} \nonumber \\
&& - \frac{\Delta}{2} \sum_{i\sigma}
\epsilon_{i} c_{i\sigma }^{\dagger }c_{i\sigma} - J \sum\limits_{i,ss^{\prime }}  \mathbf{S}_{i}c_{is}^{\dagger }
\boldsymbol{\sigma}_{ss^{\prime }} c_{is^{\prime }},
\label{ham}
\end{eqnarray}
where $c^{\dagger}_{i\sigma}$ ($c_{i\sigma}$) is the creation (annihilation) operator for electron with spin $\sigma$ at site $i$ of a honeycomb lattice. $\langle i,j \rangle$ and $\langle\langle i,j \rangle\rangle$ denote the nearest-neighbor and next-nearest-neighbor lattice sites, respectively. $t$ is the hopping parameter for the nearest-neighbor sites, and $\lambda$ is the strength of the intrinsic SOC. The sign $\nu_{ij}=\pm 1$ when the hopping direction is anti-clockwise (clockwise) \cite{KaneMele}. The honeycomb lattice is divided into two penetrating sublattices $a$ and $b$. $\epsilon_{i}=\pm 1$ when the lattice site $i$ belongs to the sublattice $a$ ($b$).
$\Delta$ is a staggered ionic potential, which breaks the sublattice symmetry of the honeycomb lattice. $\mathbf{S}_{i}$ is spin of magnetic impurity at lattice site $i$. $\boldsymbol{\sigma}=(\sigma^x,\sigma^y,\sigma^z)$ are the Pauli matrices. $J$ is the spin exchange between conduction electrons and magnetic impurities.
When the spin exchange $J=0$, the model in Eq. (\ref{ham}) is reduced to the Kane-Mele model \cite{KaneMele}. The SOC opens a gap and induces the QSH state, where the spin Hall conductivity is quantized \cite{KaneMele}. The ionic potential drives the phase transition between the QSH state and topologically trivial band insulator \cite{KaneMele}.
When
$\lambda=0$ and $\Delta=0$, Hamiltonian in Eq. (\ref{ham}) describes the double exchange of electrons
\cite{Zener}. The spin exchange drives the system from paramagnetic to magnetic states.
When the ionic potential is absent ($\Delta=0$), the spin exchange and the SOC mutually interplay, and as a consequence the model in Eq. (\ref{ham}) exhibits a topological magnetic phase transition at half filling \cite{Tien}. With a fixed SOC, the spin exchange drives the system from the topological paramagnet (QSH state) to a topological (QSH) antiferromagnet, and then to a topologically trivial antiferromagnet \cite{Tien}. We will examine the mutual effect of the ionic potential and the spin exchange on the topological magnetic phase transition at half filling. In the following we use $t=1$ as the unit of energy.

We use the dynamical mean field theory (DMFT) to calculate the Green function and its self energy \cite{Metzner,GKKR}. Within the DMFT, the self energy depends only on frequency. Therefore, the Dyson equation of the Green function reads
\begin{equation}
\hat{G}(\mathbf{k},z)=\left[ z-\hat{H}_0(\mathbf{k})-\hat{\Sigma }(z)
\right] ^{-1},
\label{gfdyson}
\end{equation}
where $\hat{G}(\mathbf{k},z)$ and $\hat{\Sigma }(z)$ is the Green function and the self energy of electrons, the creation operator of which is presented in the row vector
$(c_{\mathbf{k} a \uparrow }^\dagger, c_{\mathbf{k} b\uparrow }^\dagger, c_{\mathbf{k} a\downarrow }^\dagger, c_{\mathbf{k} b \downarrow }^\dagger)$.
$\hat{H}_0(\mathbf{k})$ is the non-interacting Bloch Hamiltonian, which has the block form in the spin index
\begin{eqnarray}
\hat{H}_0(\mathbf{k}) = \left(
\begin{array}{cc}
\hat{h}_{\uparrow }(\mathbf{k}) & 0 \\
0 & \hat{h}_{\downarrow }(\mathbf{k})
\end{array}
\right) ,
\end{eqnarray}
where
\begin{eqnarray*}
\hat{h}_{\sigma }(\mathbf{k}) =\left(
\begin{array}{cc}
 - \sigma \lambda \xi_{\mathbf{k}} - \Delta/2 & -t\gamma_{\mathbf{k}} \\
-t\gamma_{\mathbf{k}}^{\ast } & \sigma \lambda \xi_{\mathbf{k}} + \Delta/2
\end{array}%
\right) .
\end{eqnarray*}
Here we have used the notations:
$\gamma_{\mathbf{k}}=\sum_{\delta} e^{i \mathbf{k} \cdot \mathbf{r}_{\delta}}$, $\xi_{\mathbf{k}}=i\sum_{\eta }\nu _{\eta }e^{i\mathbf{k}\cdot
\mathbf{r}_{\eta }}$, where $\delta$ and $\eta$ denotes nearest-neighbor and  next-nearest-neighbor sites of a given site in the honeycomb lattice, respectively. Within the DMFT the self energy is diagonal $\hat{\Sigma}(z)=\text{diag}[\Sigma_{a\uparrow}(z), \Sigma_{b\uparrow}(z),\Sigma_{a\downarrow}(z),\Sigma_{b\downarrow}(z)]$, because the DMFT neglects intersite correlations. Therefore, the Green function also has the block form in the spin index. This allows us to consider the electron dynamics and the topology in each spin sector.
The self energy
$\Sigma_{\alpha\sigma}(z)$ is calculated from a single site of the $\alpha$-sublattice embedded in a self-consistent dynamical mean field \cite{Tien}. For classical spin of magnetic impurities, the effective single site can exactly be solved and the self energy can self consistently be determined \cite{Tien}. Without loss of generality, we set the magnitude of the impurity spin $S=1$.
Once the self energy and the Green function are obtained, we can compute the magnetization and determine the magnetic nature of the ground state. The magnetization of the $\alpha$-sublattice is determined by
\begin{eqnarray}
m_\alpha = \frac{1}{2N} \sum_{\mathbf{k},\sigma} \sigma \langle c^\dagger_{\mathbf{k}\alpha \sigma} c_{\mathbf{k}\alpha \sigma} \rangle ,
\end{eqnarray}
which can directly be calculated from the Green function in Eq. (\ref{gfdyson}).
When $m_a=m_b \neq 0$, the ground state is ferromagnetic, and when $m_a=-m_b \neq 0$ it is antiferromagnetic. The Chern number in Eq. (\ref{chern}) can also be computed when the self energy is obtained \cite{Fukui}. It determines the topological nature of the ground state.

\begin{figure}[t]
\includegraphics[width=0.49\textwidth]{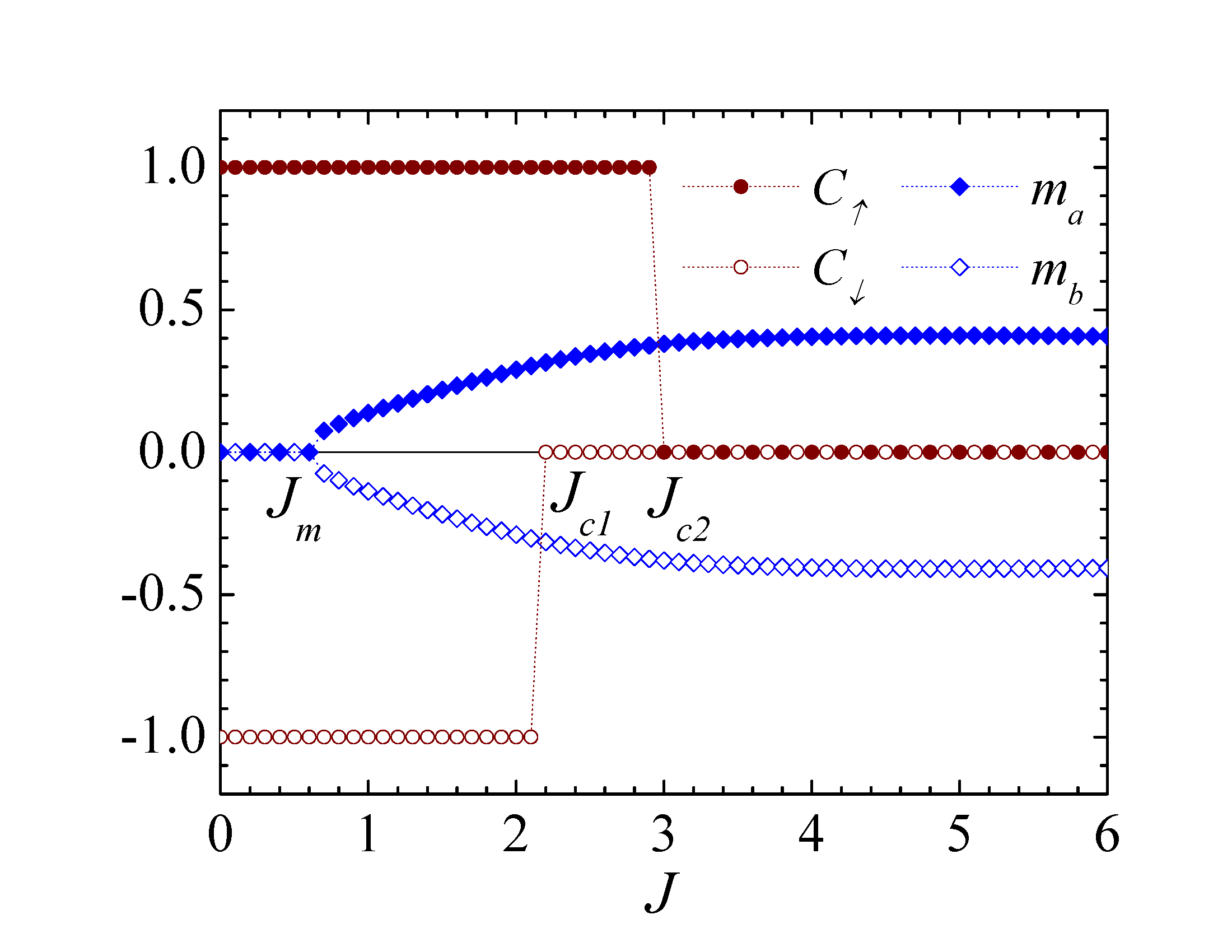}
\caption{(Color online) The Chern number $C_{1\sigma}$ for spin orientation $\sigma$ and the sublattice magnetization $m_{\alpha}$ via the spin exchange $J$ at fixed SOC $\lambda=0.5$ and ionic potential $\Delta=1$.}
\label{fig1}
\end{figure}

In Fig. \ref{fig1}  we plot the sublattice magnetizations and the Chern number for each spin orientation as functions of the spin exchange $J$ at given model parameters, where the ground state is the $Z_2$ topological insulator at $J=0$ \cite{KaneMele}. There is a critical spin exchange $J_m$ (for $\lambda=0.5$ and $\Delta=1$, $J_m \approx 0.6$), which separates the antiferromagnetic ground state ($m_a=-m_b \neq 0$) from the paramagnetic one ($m_a=m_b=0$). The magnetic phase transition is a common feature of the double exchange model \cite{Zener}. The SOC and the ionic potential do not qualitatively change the phase transition. However, the Chern number significantly changes when the ionic potential is present. Figure \ref{fig1} shows three different topological natures of the ground state which are separated by $J_{c1}$ and $J_{c2}$ (for $\lambda=0.5$ and $\Delta=1$, $J_{c1} \approx 2.2$, $J_{c2} \approx 3.0$). When $J<J_{c1}$, the Chern number $C_{1\uparrow}=-C_{1\downarrow}=1$. In this region the charge Hall conductivity vanishes, while the spin one is quantized. This yields the SQH effect. When $J_{c1} < J < J_{c2}$, electrons with different spin orientations have different topological natures.
Electrons with spin up forms a Chern topological insulator with $C_{1\uparrow}=1$, while electrons with spin down are in topologically trivial insulating state ($C_{1\downarrow}=0$). This shows a topologically breaking of the spin symmetry. Only when $\Delta \rightarrow 0$, $J_{c1}=J_{c2}$, the topological symmetry of two spin orientations is restored. It indicates an essential role of the ionic potential in the topologically breaking. However, the ionic potential alone cannot break the topological symmetry of two spin orientations, because in the absence of the spin 
\onecolumngrid\

\begin{figure*}[t]\
\hspace{0.4cm}$J=1.0$ \hspace{2.5cm} $J=2.2$ \hspace{2.3cm} $J=2.5$ \hspace{2.3cm} $J=3.0$ \hspace{2.3cm} $J=4.0$  \\
\vspace{0.2cm}
\includegraphics[width=0.19\textwidth]{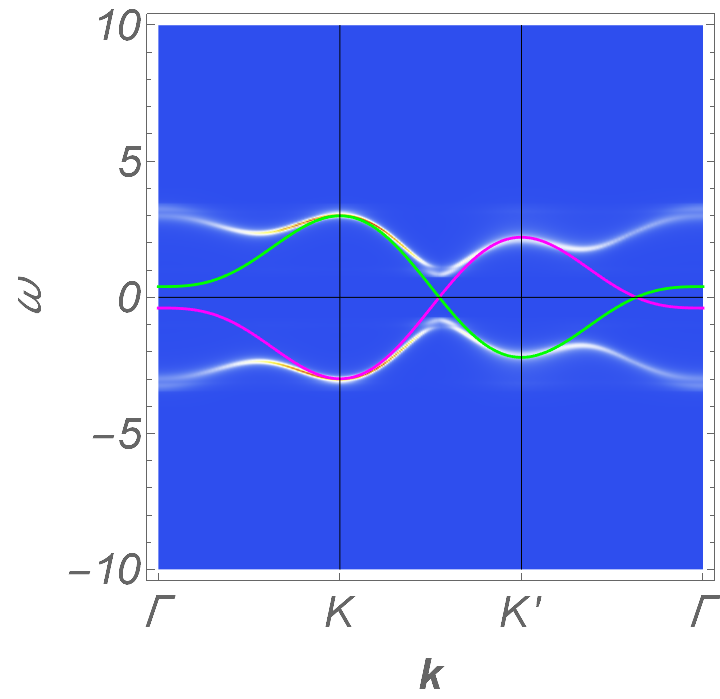} \hspace{0.cm}
\includegraphics[width=0.19\textwidth]{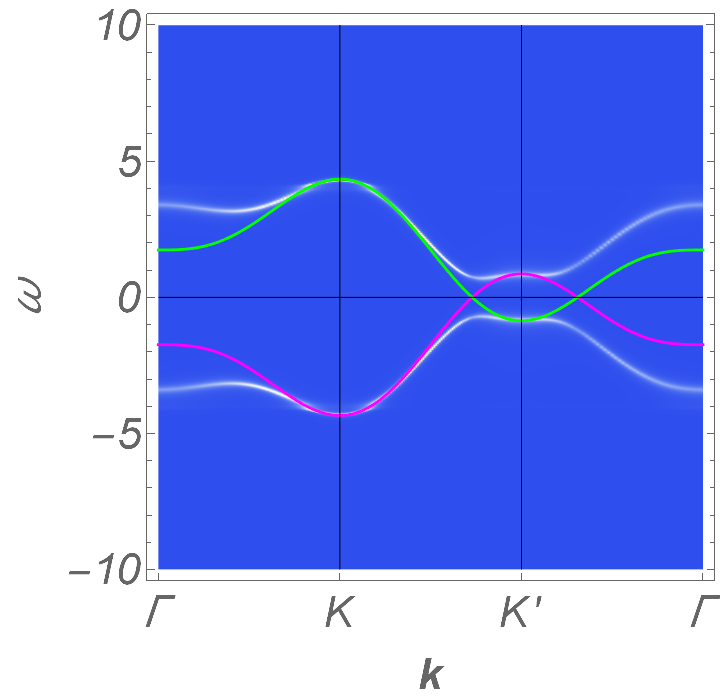} \hspace{0.cm}
\includegraphics[width=0.19\textwidth]{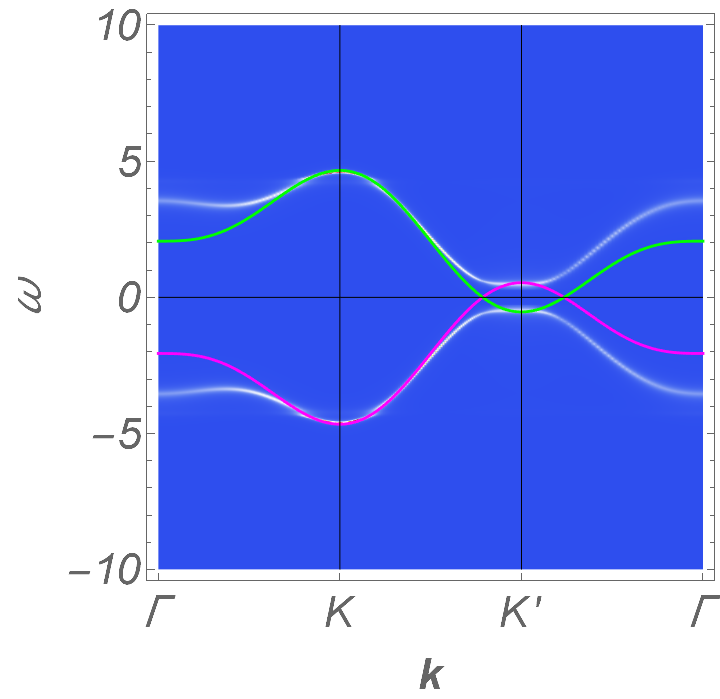} \hspace{0.cm}
\includegraphics[width=0.19\textwidth]{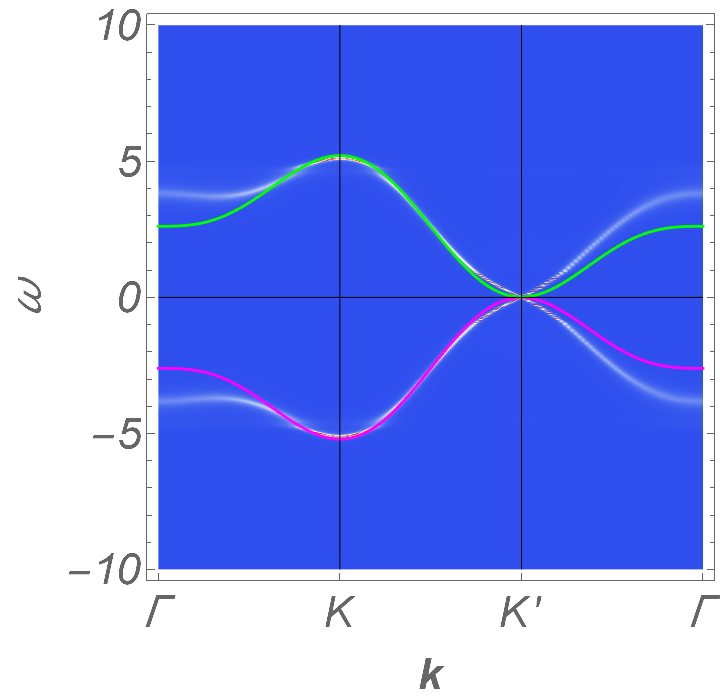} \hspace{0.cm}
\includegraphics[width=0.19\textwidth]{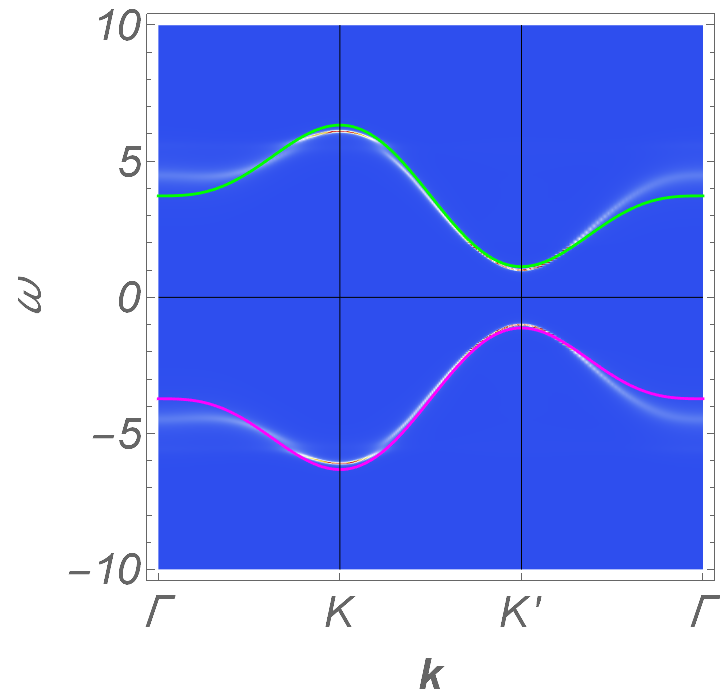} \\
\vspace{0.3cm}
\includegraphics[width=0.19\textwidth]{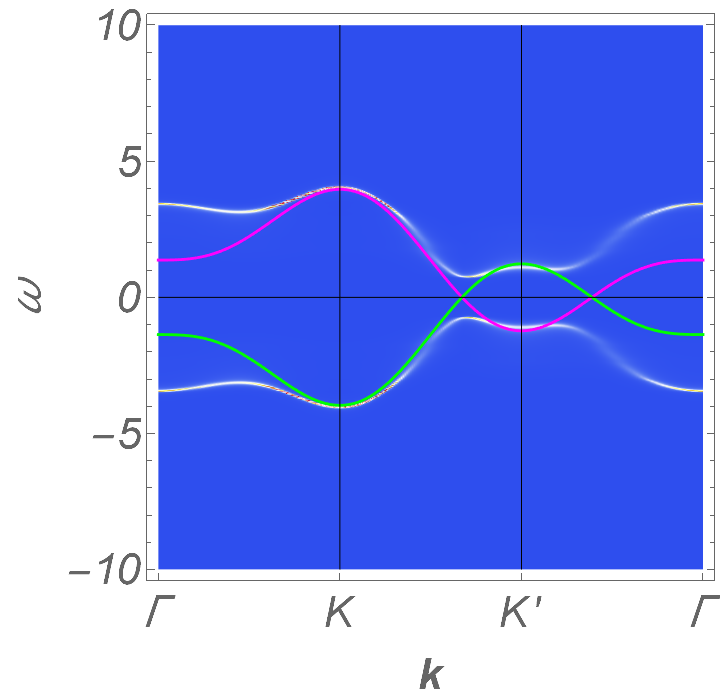} \hspace{0.cm}
\includegraphics[width=0.19\textwidth]{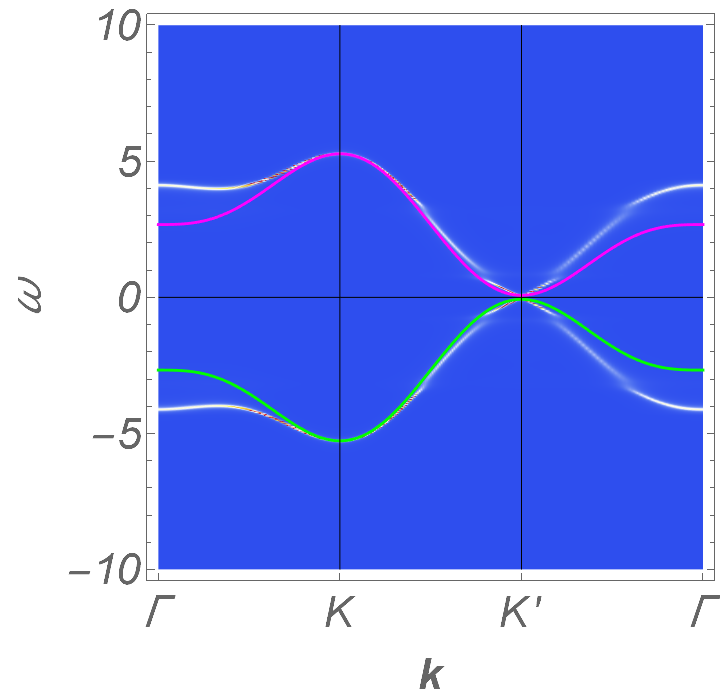} \hspace{0.cm}
\includegraphics[width=0.19\textwidth]{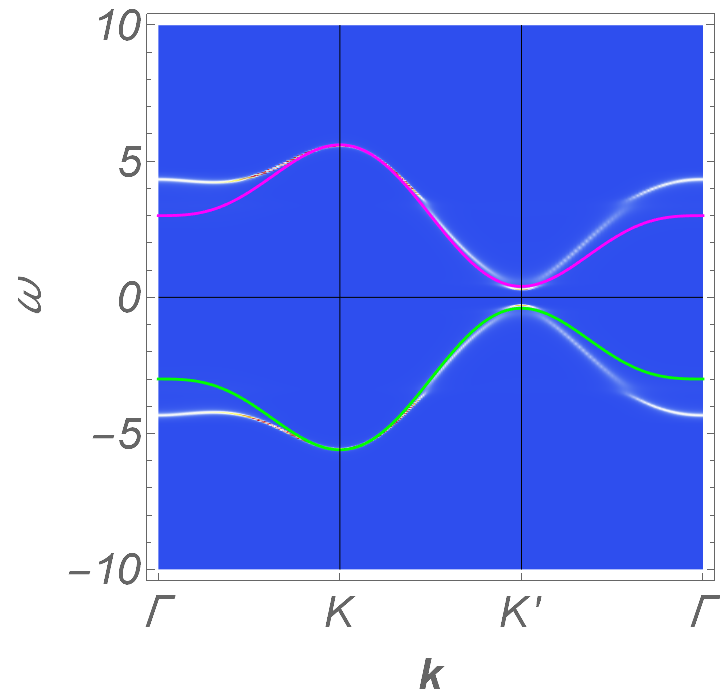} \hspace{0.cm}
\includegraphics[width=0.19\textwidth]{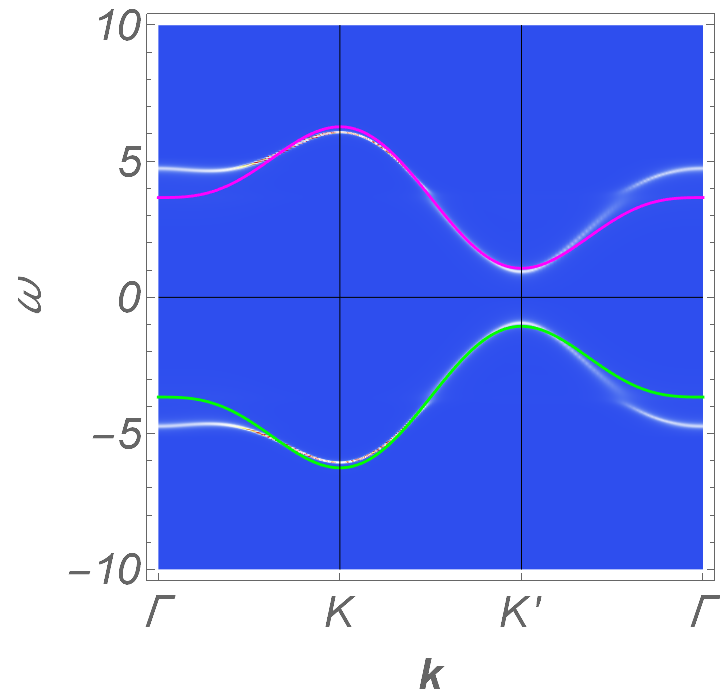} \hspace{0.cm}
\includegraphics[width=0.19\textwidth]{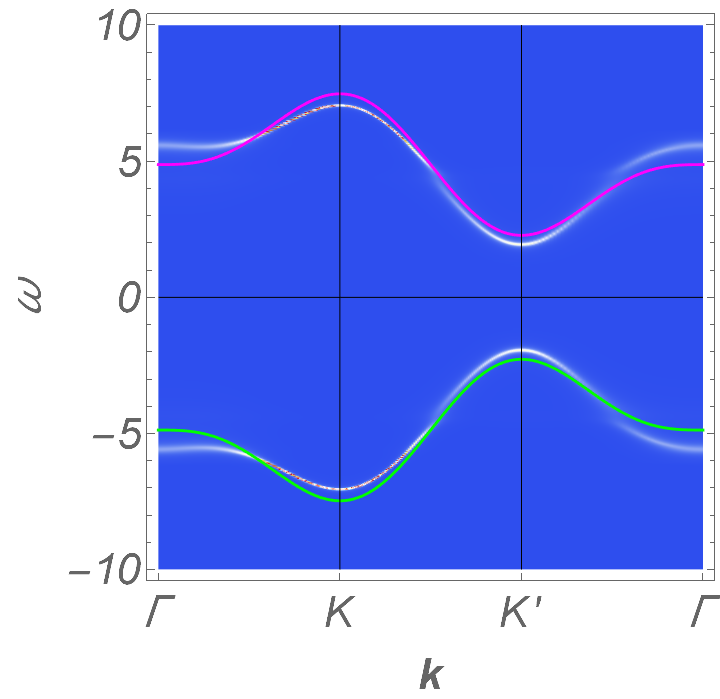} \\
\hspace{0.5cm}\includegraphics[width=0.2\textwidth]{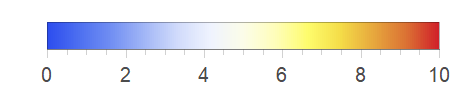} \hspace{0.5cm}
\includegraphics[width=0.08\textwidth]{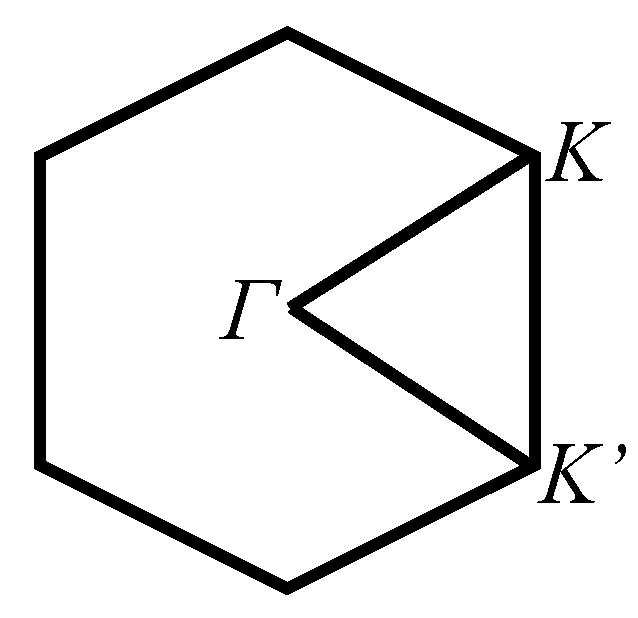}\
\caption{(Color online) The spectral function of electrons (the color density plot) and the zeros of the diagonal topological Green function (the green and magenta solid lines) along the high symmetry lines of the Brillouin zone for various values of the spin exchange $J$ and fixed SOC $\lambda=0.5$ and $\Delta=1$. The upper (lower) row presents the plots for the spin up (down) orientation.
Due to the overlap between the spectral functions of two sublattices, the color density plots show
the maximum values between them.}\
\label{fig2}\
\end{figure*}\
\twocolumngrid
\noindent
exchange, two spin orientations are topologically symmetric \cite{KaneMele}. Electrons with both spin orientations simultaneously form either topological
or topologically trivial insulators.
When $J>J_{c2}$, electrons with both spin orientations are in topologically trivial insulator, because $C_{1\uparrow}=C_{1\downarrow}=0$.

The topological nature of the ground state can also be identified by the cross of the zeros of the diagonal topological Green function. In contrast to the TKNN formula, the zeros are easily determined. We obtain the analytical expression of the zeros
\begin{eqnarray}
\mathcal{E}_{\sigma}^{(\alpha)}(\mathbf{k})= - \epsilon_{\bar{\alpha}}\sigma \lambda \xi_{\mathbf{k}} - \epsilon_{\bar{\alpha}} \Delta/2 + \Sigma_{\bar{\alpha}\sigma}(i0),
\label{zeros}
\end{eqnarray}
where $\bar{\alpha}$ denotes the sublattice other than the $\alpha$ sublattice.
In Fig. \ref{fig2} we plot the zeros of the diagonal topological Green function and the spectral function
$\rho_{a\sigma}(\mathbf{k},\omega)=-\text{Im} G_{\alpha\sigma}(\mathbf{k},\omega+i0^+)/\pi$ in three typical topological phase regions.
The peaks of the spectral function resemble the quasiparticles and describe the energy bands. Figure \ref{fig2} shows that
the spectral functions clearly display upper and lower bands separated by a gap for all values of $J$.
Both the SOC and the spin exchange open the gap \cite{Tien}. However, the spin exchange first reduces the band gap opened by the SOC, and then increases the band gap after closing it \cite{Tien}. The gapless points are the evidence of the topological phase transition.
In the region of weak spin exchange $J<J_{c1}$, in both spin sectors the zeros
cross each other as shown in Fig. \ref{fig2}. The zero's cross is the evidence of a topological state, because it is a consequence of the coincidence of the poles and zeros, or equivalently, of the vortex existence.
Therefore, electrons with both spin orientations in the topological insulating state. However, the zero's cross alone cannot reveal the value of the Chern number. Because this region adiabatically connects with the $Z_2$ topological insulator at $J=0$, the ground state could be the QSH one.
In the intermediate region $J_{c1} < J < J_{c2}$, in the spin-up sector, the zeros cross each other, while in the spin-down sector they do not cross. This indicates that electrons with spin up form a Chern topological insulator, and electrons with spin down are in a topologically trivial insulating
state. The cross behavior of the zeros is consistent with
the topological nature identified by the Chern number. When the spin exchange is large, $J > J_{c2}$,
in both spin sectors the zeros do not cross each other as shown in Fig. \ref{fig2}. This shows that electrons with both spin orientations form topologically trivial insulator. In contrast to the TKNN formula, the topological nature is simpler analyzed through the zero's behavior. Equation (\ref{zeros}) clearly shows that the spin exchange impacts on the zero's behavior through the self energy at zero frequency.
In the antiferromagnetic state $\Sigma_{a\uparrow}(i0)-\Sigma_{a\downarrow}(i0)=-[\Sigma_{b\uparrow}(i0)-\Sigma_{b\downarrow}(i0)]$, and the ionic potential plays like a staggered chemical potential, which
leads to  $\Sigma_{a\uparrow}(i0)+\Sigma_{a\downarrow}(i0)=-[\Sigma_{b\uparrow}(i0)+\Sigma_{b\downarrow}(i0)]$. Therefore, the self energy can be presented as
\begin{eqnarray}
\Sigma_{\alpha\sigma}(i0) = \epsilon_{\alpha} M + \epsilon_{\alpha} \sigma \delta M ,
\label{se0}
\end{eqnarray}
where $M=[\Sigma_{a\uparrow}(i0)+\Sigma_{a\downarrow}(i0)]/2$, and $\delta M=[\Sigma_{a\uparrow}(i0)-\Sigma_{a\downarrow}(i0)]/2$.
Equation (\ref{se0}) shows that the self energy $\Sigma_{\alpha\sigma}(i0)$ plays like a combination of staggered non-magnetic and magnetic fields.
Therefore, in different spin sectors the spin exchange shifts the zeros apart differently. As a consequence, in the intermediate region, the zeros in the spin-down sector do not cross each other, while the zeros in the spin-up sector still cross. This leads to a topologically breaking of the spin orientations. When the spin exchange is absent ($J=0$), the self energy vanishes, and the ionic potential only describes the phase transition from topological insulator to topologically trivial insulator \cite{KaneMele}.
When the ionic potential vanishes ($\Delta=0$), $M=0$, the topological symmetry of the spin orientations is restored \cite{Tien}.
The limiting cases show that neither the spin exchange nor the ionic potential alone can break the topological symmetry of two spin orientations.
The topologically breaking of the spin symmetry is a mutual effect of the spin exchange and the ionic potential.
It is a challenge for observing such topologically breaking of the spin symmetry by experiments.
Indeed, the Haldane model was experimentally realized by ultracold atoms \cite{Jotzu}. One may expect the Kane-Mele model or the spin version of the Haldane model may be realized too. When a staggered field like the self energy in Eq. (\ref{se0}) is imposed over the quantum simulated lattice, the  topologically breaking of the spin symmetry would be
observed.
When the spin exchange is large $J>J_{c2}$, the self energy $\Sigma_{\alpha\sigma}(i0)$ is large enough for shifting the zeros in both spin sectors away from each other. Therefore electrons with both spin orientations are in topologically trivial insulator.

At the topological phase transition point $J_{c1}$ ($J_{c2}$), the spectral functions of the sublattice $a$ and $b$ touch each other, as shown in Fig. \ref{fig2}. This shows a gapless state in the spin down (up) sector. The occurrence of the gapless state is a consequence of the change of the topological invariant at the phase transition point, and it is an evidence of the bulk-edge correspondence. Figure \ref{fig2} also shows a touch of the zeros at the topological phase transition. Although the Cauchy interlacing inequalities in Eq. (\ref{interlacing}) were proved only for the topological Green function, the zero's touch at the gapless points shows that the Cauchy interlacing inequalities may also be valid at the gapless points for the poles of the full interacting Green function and the zeros of the diagonal topological Green function.
It also shows that at the gapless points, the energy bands of interacting fermions, the poles and the zeros of the diagonal topological Green function touch each other. We can use either the energy bands, the poles or the zeros to determine the gapless points, such as the Weyl nodes of the Weyl semimetals \cite{Knap}.
However, except for the gapless points, the zeros of the diagonal topological Green function do not always lie within the band gap, especially when the spin exchange is strong.

\section{Conclusion}

We have constructed the topological Green function, which can describe the topological nature of interacting fermions.
In two-dimensional insulators, the Hall conductivity calculated by the topological Green function is equal to the one of the original interacting fermions. Based on the eigenvector-eigenvalue identity, it was shown that when the zeros and the poles of the diagonal topological Green function coincide, the vortices and their nonzero charge exist.  The cross of the zeros of the diagonal topological Green function in the momentum space is a simple and useful tool to determine the topological nature of the ground state. As an application, we have identified the topological nature of a modeled magnetic insulator. It is found that the interplay between the spin exchange and ionic potential can lead to the topologically breaking of the spin symmetry. In the antiferromagnetic state, electrons with one spin orientation form the topological insulator, while electrons with the opposite spin orientation are in topologically trivial one. The topological nature identification by the zero's cross is consistent with the topological invariant of interacting electrons. However, the zero's cross can only identify the topological nature and it seems that it cannot reveal
the non-zero value of the topological invariant.
The topological Green function can also describe the topology of gapless systems. The gapless points can be determined by the zero's touch in the momentum space.
In the non-interacting systems, the degeneracy of the zeros corresponds
to the degeneracy of the
edge states \cite{Yamaji}.
In interacting systems it is not clear that the zero-edge correspondence is still valid, because the topological Green function differs from the full Green function of interacting particles. We leave the problem for further study.

\section*{Acknowledgement}

This research is funded by Vietnam National Foundation for Science and Technology Development (NAFOSTED) under Grant No 103.01-2019.309.


\begin{thebibliography}{99}

\bibitem{Assaad}
M. Hohenadler and F. F. Assaad, J. Phys.: Condens. Matter \textbf{25},  143201 (2013).

\bibitem{Rachel}
S. Rachel, Rep. Prog. Phys. \textbf{81},  116501 (2018).

\bibitem{Abrikosov}
A. A. Abrikosov, L. P. Gorkov, and I. E. Dzyaloshinski, {\em Method of Quantum Field Theory in Statistical Physics} (Dover, New York, 1975).

\bibitem{Volovik}
G. E. Volovik, Zh. Eksp. Teor. Fiz. \textbf{94}, 123 (1988) [Sov. Phys. JETP
67, 1804 (1988)].

\bibitem{Wang1}
Z. Wang, X.-L. Qi, and S.-C. Zhang,
Phys. Rev. Lett. \textbf{105}, 256803 (2010).

\bibitem{Xie}
L. Wang, X. Dai, and X. C. Xie,
Phys. Rev. B \textbf{84}, 205116 (2011).

\bibitem{Xie1}
L. Wang, H. Jiang, X. Dai, and X. C. Xie,
Phys. Rev. B \textbf{85}, 235135 (2012).

\bibitem{Gurarie}
V. Gurarie, Phys. Rev. B \textbf{83}, 085426 (2011).

\bibitem{Wang}
Z. Wang and S.-C. Zhang, Phys. Rev. X \textbf{2}, 031008 (2012).

\bibitem{Wang2}
Z. Wang and S.-C. Zhang,
Phys. Rev. B \textbf{86}, 165116 (2012).

\bibitem{Wang3}
Z. Wang and B. Yan, J. Phys. Condens. Matter \textbf{25}, 155601 (2013).

\bibitem{Tien4}
Minh-Tien Tran, T. Takimoto, and K.-S. Kim,
Phys. Rev. B \textbf{85}, 125128 (2012).

\bibitem{Son}
Hong-Son Nguyen and Minh-Tien Tran,
Phys. Rev. B \textbf{88}, 165132 (2013).

\bibitem{Tien}
Minh-Tien Tran, Hong-Son Nguyen and Duc-Anh Le, Phys. Rev. B \textbf{93}, 155160 (2016).

\bibitem{Mai}
Thanh-Mai Thi Tran, Duc-Anh Le, Tuan-Minh Pham, Kim-Thanh Thi Nguyen, and Minh-Tien Tran,
Phys. Rev. B \textbf{102}, 205124 (2020).


\bibitem{Kawakami}
R. Peters, T. Yoshida, and N. Kawakami
Phys. Rev. B \textbf{98}, 075104 (2018).

\bibitem{Valenti}
T. Mertz, K. Zantout, and R. Valenti,
Phys. Rev. B \textbf{100}, 125111 (2019).


\bibitem{Knap}
W. Witczak-Krempa, M. Knap, and D. Abanin,
Phys. Rev. Lett. \textbf{113}, 136402 (2014).

\bibitem{Hatsugai0}
Y. Hatsugai, J. Phys.: Condens. Matter \textbf{9}, 2507 (1997).

\bibitem{Hatsugai}
Y. Morita and Y. Hatsugai,
Phys. Rev. B \textbf{62}, 99 (2000).

\bibitem{Denton}
P. B. Denton, S. J. Parke, T. Tao and X. Zhang,
preprint arXiv:1908.03795 (2019).

\bibitem{Yamaji}
T. Misawa and Y. Yamaji, preprint arXiv:2102.04665 (2021).

\bibitem{Balents}
R.-J. Slager, L. Rademaker, J. Zaanen, and L. Balents,
Phys. Rev. B \textbf{92}, 085126 (2015).

\bibitem{Weng}
H. Weng, R. Yu, X. Hu, X. Dai,  and Z. Fang,
Adv. Phys. \textbf{64}, 227 (2015).

\bibitem{Tokura}
Y. Tokura, K. Yasuda, and A. Tsukazaki,
Nat. Rev. Phys. \textbf{1}, 126 (2019).

\bibitem{KaneMele}
C. L. Kane and E. J. Mele,
Phys. Rev. Lett. \textbf{95}, 146802 (2005).

\bibitem{Zener}
C. Zener, Phys. Rev. \textbf{82}, 403 (1951).

\bibitem{Hedin}
L. Hedin and S. Lundqvist, in {\it Solid State Physics}, edited by F. Seitz, D. Turnbull, and  H. Ehrenreich (Academic Press, New York, 1969), Vol. \textbf{23}, p.1.

\bibitem{TKNN}
D. J. Thouless, M. Kohmoto, M. P. Nightingale, and M. den Nijs,
Phys. Rev. Lett. \textbf{49} 405 (1982).

\bibitem{Hofstetter}
J.-H. Zheng and W. Hofstetter,
Phys. Rev. B \textbf{97}, 195434 (2018).

\bibitem{Metzner}
W. Metzner and D. Vollhardt, Phys. Rev. Lett. \textbf{62}, 324
(1989).

\bibitem{GKKR}
A. Georges, G. Kotliar, W. Krauth, and M. J. Rozenberg,
Rev. Mod. Phys. \textbf{68}, 13 (1996).


\bibitem{Fukui}
T. Fukui, Y. Hatsugai, and H. Suzuki, J. Phys. Soc. Jpn. \textbf{74}, 1674 (2005).

\bibitem{Jotzu}
G. Jotzu, M. Messer, R. Desbuquois, M. Lebrat, T. Uehlinger, D. Greif, and T. Esslinger,
Nature (London) \textbf{515}, 237 (2014).

\end{thebibliography}
\end{document}